\documentclass[12pt]{iopart}
\usepackage{epsfig}
\begin{document}
\jl{4}
\title{Group theory and the Pentaquark}
\author{Brian G Wybourne\footnote[1]{E-mail: bgw@phys.uni.torun.pl
\hskip5mm WEB: www.phys.uni.torun.pl/$\sim$bgw}}
\address{ Instytut Fizyki, Uniwersytet Miko\l{}aja Kopernika,
ul. Grudzi\c{a}dzka 5/7, 87-100 Toru\'n, Poland}
\begin{abstract}
The group classification of exotic pentaquarks involving four standard quarks
and a single antiquark to produce strangeness ${\cal S} = +1$ is outlined.
\end{abstract}

\def\Irrep{Irreducible\ representation}

\def\w{\omega}
\def\n{\noindent}
\def\h{{\textstyle{{1\over2}}}}
\def\->{\rightarrow}
\def\irrep{irreducible\ representation}
\def\t{\times}
\overfullrule=0pt
\def\'{\prime}
\def\bm{\bordermatrix}
\def\e{\varepsilon}
\def\u{{\textstyle{{2\over3}}}}
\def\d{{\textstyle{{1\over3}}}}
\def\4{{\textstyle{{4\over3}}}}
\def\3{{\textstyle{{3\over2}}}}
\def\ub{\bar{u}}
\def\db{\bar{d}}
\def\sb{\bar{s}}
\def\S{${\cal S}$}
\centerline{\today}
 
\vskip5mm

\vfill\eject
\section{Introduction}
. The possibility of observable exotic multiquark states has been the subject 
of speculation for over three decades and has generated a large 
literature, a miniscule sample might include\cite{Chodos,Jaffe,Strott,AG,Wy}. The MIT
bag model played a central role in many calculations\cite{Chodos}. Searches for
exotic multiquark resonances have been largely unsuccessful in spite of efforts
over three decades. Among the possible resonances have been those 
associated with the pentaquarks involving four quarks and a single 
antiquark, their existence having been speculated more than three decades 
ago\cite{Strott,AThesis,BW}. Evidence for a pentaquark with strangeness 
\S = +1 has been recently reported by three independent experimental 
groups\cite{Nak,Diana,Step} involving three different types of experiments..

Herein we discuss the group classification of the
states of pentaquarks. Throughout we label \irrep{s} of the relevant $SU_n$ 
groups by appropriate
parti-tions\cite{GRE,MacD}. The usual $u,d,s$ quarks have 
spin $S = \h$ with colour corresponding to the $\{1\}^c$ \irrep\ of the group
$SU_3^c$ and flavour corresponding to the $\{1\}^f$
\irrep\ of $SU_3^{fl}$. The complete set of the basic quarks transform as
the vector \irrep\ $\{1\}$ of the group $SU_{18}$\cite{Wy}. A system of 
$N$ quarks will span the totally antisymmetric \irrep\ $\{1^N\}$ of $SU_{18}$ 
which may be decomposed under the group chain
$$SU_{18}^q \-> SU_2^S \t SU_3^c \t SU_3^{fl}\eqno(1)$$
A system of $N^{\'}$ antiquarks $\bar{q}$ will span the antisymmetric 
conjugate \irrep\ $\{\overline{1^{N^{\'}}}\}$ of another group 
$SU_{18}^{\bar{q}}$ which may be decomposed under the group chain
$$SU_{18}^{\bar{q}} \-> SU_2^S \t SU_3^c \t SU_3^{fl}\eqno(2)$$  
The totally antisymmetric states formed by configurations of $N$ quarks and 
$N^{\'}$ antiquarks will span the \irrep\ $\{1^{N+N^{\'}}\}$ of the group
$SU_{36}$ leading to the classification group chain
$$SU_{36} \-> SU_{18}^q \t SU_{18}^{\bar{q}}\eqno(3)$$
Hereon we examine the various group decompositions relevant to the pentaquark.
We start with the group $SU_{36}$ and systematically reduce the symmetry
using the appropriate group decomposition branching rules until finally
the quantum numbers labelling the individual pentaquark are determined.
It is demonstrated that there is a single pentaquark state of strangeness
\S = +1 of charge $q = +1$ arising from the combination of a sextet multiplet
of quarks and a single strange antiquark. These states form part of a 
$SU_3^{fl}$ octet, $\{21\}$, and antidecuplet, $\{3^2\}$, with the relevant 
pentaquark at the apex of the antidecuplet. Other pentaquarks of
strangeness \S = +1 arise but with uniquely distinguishable quantum numbers.
\section{The $SU_{36} \-> SU_{18}^q \t SU_{18}^{\bar{q}}$ decompositions}
Assuming that under $SU_{36} \-> SU_{18}^q \t SU_{18}^{\bar{q}}$
$$\{1\} \-> \{1\}^q + \{\bar1\}^{\bar{q}}\eqno(4)$$
we have for $k$ particles
$$\eqalignno{
\{1^k\}& \-> (\{1\}^q + \{\bar1\}^{\bar{q}})\otimes\{1^k\}\cr
   & = \sum_{x=0}^k\{1^{k-x}\}^q\times\{{\bar1}^x\}^{\bar{q}}&(5)\cr}
$$
where we use $\otimes$ to indicate the operation of 
plethysm\cite{GRE,MacD,Lit,Wyb1} 
\section{The colour singlet pentaquarks}
For $4$ quarks and a single antiquark we have just the 
$SU_{18}^q \t SU_{18}^{\bar{q}}$ \irrep\ $\{1^4\}^q\times\{\bar1\}^{\bar{q}}$
to consider. We first note that for $SU_3^c\times SU_3^{fl}$\cite{King}
$$(\{1\}^c\times\{1\}^{fl})\otimes\{\rho\} = \sum_{\e\vdash \w_\rho}
\{\e\}^c\times\{\e\circ\rho\}^{fl}\eqno(6)$$

Thus under $SU_{18}^q \-> SU_2^S \t SU_3^c \t SU_3^{fl}$ we have
$$\eqalignno{
\{1^4\}^q& \-> (\{1\}^S\t\{1\}^c\t\{1\}^{fl})\otimes\{1^4\}\cr
 & = \{4\}\t(\{1\}^c\t\{1\}^{fl})\otimes\{1^4\}\cr
 &\  + \{31\}\t(\{1\}^c\t\{1\}^{fl})\otimes\{21^2\}\cr
 &\ + \{2^2\}\t(\{1\}^c\t\{1\}^{fl})\otimes\{2^2\}&(7)\cr}
$$
Making repeated use of (6) and replacing the \irrep{s} of $SU_2^S$ by the
appropriate spin multiplicities ($2S+1$) as a left superscript, dropping the
colour (c) and flavour (fl) superscripts, we have for
the complete decomposition
$$\eqalignno{
\{1^4\} &\-> {}^5(\{31\}\{1\} + \{1\}\{31\} +\{2^2\}\{2^2\})\cr
        & + {}^3(\{4\}\{21^2\} + \{31\}\{31\} + \{31\}\{21^2\} + \{2^2\}\{31\}
+\{2^2\}\{1\}\cr
& + \{1\}\{4\} + \{1\}\{2^2\} + \{1\}\{21^2\})\cr
& + {}^1(\{4\}\{2^2\} + \{31\}\{31\} + \{31\}\{21^2\} + \{2^2\}\{4\}
+ \{2^2\}\{2^2\}\cr
& + \{1\}\{31\} + \{1\}\{21^2\})&(8)\cr}
$$
The single antiquark transforms as the $\{\bar1\}$ \irrep\ of 
$SU_{18}^{\bar{q}}$ and under (2) decomposes as
$$\{\bar1\} \-> {}^2\{\bar1\}^c\t\{\bar1\}^{fl}\eqno(9)$$
To form pentaquark states with $4$ quarks and an antiquark we must form the 
product of the $4$ quark states arising from (8) with those of (9). Under the
usual postulates of QCD only colourless states correspond to observables. Thus
we seek the pentaquark states that transform under the group $SU_3^c$ as the
$\{0\}^c$ \irrep. This is only possible for those four quark states that
transform as $\{1\}^c$ under $SU_3^c$. Note that under $SU_3^c$ 
$$\{1\}^c\t\{\bar1\}^c = \{0\}^c + \{21\}^c\eqno(10)$$
Thus from (8) the only possible pentaquark states will come from
$$\hskip-1cm^5(\{1\}\{31\}) + {}^3(\{1\}\{4\} + \{1\}\{2^2\} + \{1\}\{21^2\})
 + {}^1(\{1\}\{31\} + \{1\}\{21^2\})\eqno(11)$$
Recall the $SU_3$ Kronecker products
$$\eqalignno{
\{21^2\}\t\{\bar1\} & = \{0\} + \{21\}&(12a)\cr
\{2^2\}\t\{\bar1\} & = \{21\} + \{3^2\}&(12b)\cr
\{31\}\t\{\bar1\} & = \{21\} + \{3\} + \{42\}&(12c)\cr
\{4\}\t\{\bar1\} & = \{3\} + \{51\}&(12d)\cr}
$$
Combining these with (9) and keeping only colour singlets gives the possible
pentaquark states under $SU_2^S\t SU_3^{fl}$ as
$$\eqalignno{
&{}^{6,4}(\{21\} + \{3\} + \{42\})&(13a)\cr
&{}^{4,2}(\{0\} + 2\{21\} + \{3\} + \{3^2\} + \{51\})&(13b)\cr
&{}^2(\{0\} + 2\{21\} + \{3\} + \{42\})&(13c)\cr}
$$ 
where again the spin multiplicities appear as left superscripts.

\section{Quarks and the $SU_3^{fl} \-> U_1^Y\t SU_2^I$ decompositions}
The standard charge ($q$), isospin projection  ($I_z$), strangeness (\S)
and hypercharge ($Y$) of the quark triplet are tabulated below

\begin{center}
\begin{tabular}{ccccc}
quark&$q$&$I_z$&\S&$Y$\\
$u$&$\u$&$\h$&$0$&$\d$\\
$d$&$-\d$&$-\h$&$0$&$\d$\\
$s$&$-\d$&$0$&$-1$&$-\u$\\
\end{tabular}
\end{center}

Under $SU_3^{fl} \-> U_1^Y\t SU_2^I$ the triplet \irrep\ of $SU_3^{fl}$
decomposes as
$$\{1\}^{fl} \-> \{\d\}\t\{1\} + \{-\u\}\t\{0\}\eqno(14)$$
For an arbitrary \irrep\ $\{\lambda\}^{fl}$ of $SU_3^{fl}$ we have the 
decompositon
$$\eqalignno{
\{\lambda\}^{fl}& \-> ( \{\d\}\t\{1\} + \{-\u\}\t\{0\})\otimes\{\lambda\}\cr
& = \sum_{\rho}( \{\d\}\t\{1\})\otimes\{\lambda/\rho\}\cdot
(\{-\u\}\t\{0\})\otimes\{\rho\}&(15)\cr}
$$
Only the leading term in (15) 
$$\eqalignno{
( \{\d\}\t\{1\})\otimes\{\lambda\}& = \sum_{\e}\{\d\}\otimes\{\e\}\cdot
\{1\}\otimes\{\e\circ\lambda\}\cr
& = \{{{\w_\lambda}\over3}\}\t\{\lambda\},&(16)\cr}
$$
where the right-hand-side of (16) is an \irrep\ of $U_1^Y\t SU_2^I$,
can yield non-strange $4$ quark states. Furthermore, for non-strange $4$ quark
states it is necessary that the the number of parts of the partition 
$(\lambda)$ have less than three parts. Thus the \irrep\ 
$\{21^2\} \equiv \{1\}$ appearing in (8),(11) and (12) cannot yield 
non-strange $4$ quark states. Noting (11) and 
using (16) we have the relevant non-strange leading terms as below

\begin{center}
\begin{tabular}{ccccccccccc}
$SU_3^{fl}$&$U_1^Y\t SU_2^I$&$I$\\
$\{2^2\}$&$\{\4\}\t\{0\}$&$0$&&&&&&&(17a)\\
$\{31\}$&$\{\4\}\t\{2\}$&$1$&&&&&&&(17b)\\
$\{4\}$&$\{\4\}\t\{4\}$&$2$&&&&&&&(17c)\\
\end{tabular}
\end{center}

The states associated with the maximal isospin projection in (17a,b,c,d) will
be respectively

$$\eqalignno{
&uudd&(18a)\cr
&uuud&(18b)\cr
&uuuu&(18c)\cr}
$$
Pentaquarks of strangeness \S = +1 may be formed by combining each member of
each isospin multiplet with a single strange antiquark. The resulting
pentaquarks, along with their hypercharge ($Y$), isospin projection ($I_z$)
and electric charge $(q)$ are given in Table 1. 
\begin{center}
\begin{tabular}{ccccc}
&$I$&$Y$&$I_z$&$q$\\
$\{4\}\sb$&$2$&$2$&&\\
$uuuu\sb$&&&$2$&$+3$\\
$uuud\sb$&&&$1$&$+2$\\
$uudd\sb$&&&$2$&$+1$\\
$uddd\sb$&&&$-1$&$0$\\
$dddd\sb$&&&$-2$&$-1$\\
$\{31\}\sb$&$1$&$2$&&\\
$uuud\sb$&&&$1$&$+2$\\
$uudd\sb$&&&$0$&$+1$\\
$uddd\sb$&&&$-1$&$0$\\
$\{2^2\}\sb$&$0$&$2$&&\\
$uudd\sb$&&&$0$&$+1$\\
\end{tabular}
\end{center}
\noindent{\bf Table 1.} The strangeness \S = +1 pentaquarks.

To continue let us look in more detail at the $18$ pentaquark states arising 
from the product
$$\{2^2\}\t\{\bar1\} = \{21\} + \{3^2\}\eqno(19)$$
Under $SU_3 \-> U_1^Y \t SU_2^I$ we have from (15)
$$\{2^2\} \-> \{\4\}\t\{0\} + \{\d\}\t\{1\} + \{-\u\}\t\{2\}\eqno(20)$$
This gives a set of $6$ states with the following $(Y,q,I,I_z)$ numbers

\begin{center}
\begin{tabular}{ccccc}
quarks&$Y$&$q$&$I$&$I_z$\\
uudd&$\4$&$\u$&$0$&$0$\\
uuds&$\d$&$\u$&$\h$&$\h$\\
udds&$\d$&$-\d$&$\h$&$-\h$\\
uuss&$-\u$&$\u$&$1$&$1$\\
udss&$-\u$&$-\d$&$1$&$0$\\
ddss&$-\u$&$-\4$&$1$&$-1$\\
\end{tabular}
\end{center}
The six 4-quark states associated with the $\{2^2\}^{fl}$ sextet are shown in 
Figure 1.

\begin{figure}[htb]
\begin{center}
\epsfig{file=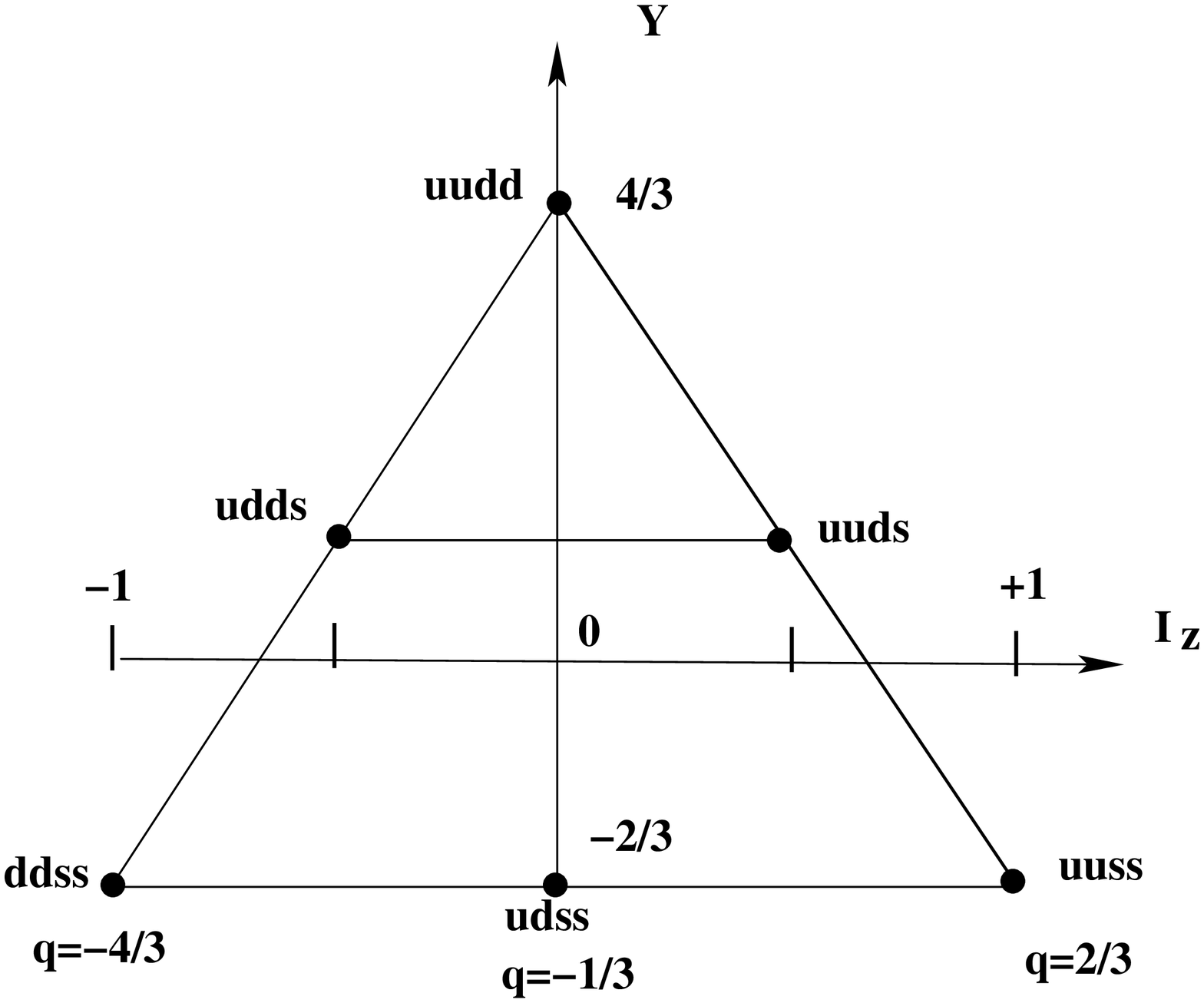,height=8cm}
\end{center}
\caption{The $\{2^2\}^{fl}$ sextet}
\end{figure}

These $6$ states, each involving $4$ quarks, may be combined with the $3$
antiquarks $(\ub,\db,\sb$ to form the $18$ pentaquark states
associated with the octet and antidecuplet arising in (19). We list below these
$18$ pentaquarks

\begin{center}
\begin{tabular}{ccccc}
penta&$I_z$&$Y$&$q$&\S\\
uudd$\ub$&$-\h$&$1$&$0$&$0$\\
uudd$\db$&$\h$&$1$&$1$&$0$\\
uudd$\sb$&$0$&$2$&$1$&$+1$\\
uuds$\ub$&$0$&$0$&$0$&$-1$\\
uuds$\db$&$1$&$0$&$1$&$-1$\\
uuds$\sb$&$\h$&$1$&$1$&$0$\\
uuss$\ub$&$\h$&$-1$&$0$&$-2$\\
uuss$\db$&$\3$&$-1$&$1$&$-2$\\
uuss$\sb$&$1$&$0$&$1$&$-1$\\
udss$\ub$&$-\h$&$-1$&$-1$&$-2$\\
udss$\db$&$\h$&$-1$&$0$&$-2$\\
udss$\sb$&$0$&$0$&$0$&$-1$\\
udds$\ub$&$-1$&$0$&$-1$&$-1$\\
udds$\db$&$0$&$0$&$0$&$-1$\\
udds$\sb$&$-\h$&$1$&$0$&$0$\\
ddss$\ub$&$-\3$&$-1$&$-2$&$-2$\\
ddss$\db$&$-\h$&$-1$&$-1$&$-2$\\
ddss$\sb$&$-1$&$0$&$-1$&$-1$\\
\end{tabular}
\end{center}

If we make a $I_z$ versus $Y$ plot of the above entries we obtain, as
expected, an antidecuplet $\{3^2\}$ on which is superimposed an octet $\{21\}$
with the reported pentaquark state at the apex of the antidecuplet. These 
states occur with spin $S = \h$ and $S = \3$.

\begin{figure}[htb]
\begin{center}
\epsfig{file=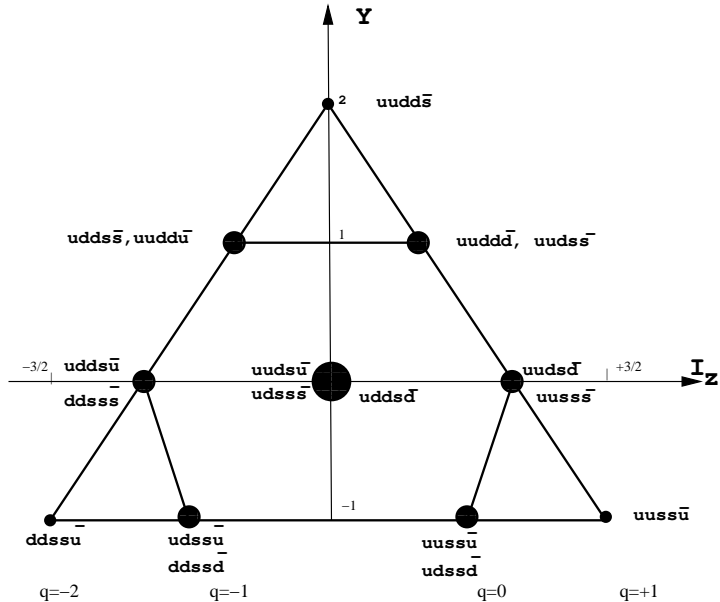,height=8cm}
\end{center}
\caption{The octet + antidecuplet}
\end{figure}
\section{Concluding remarks}
We have shown how the quantum numbers associated with states involving 
$4$ quarks and an antiquark can be readily determined. Nine pentaquarks of
strangeness \S = +1 and spin $S = \h$ are found. They may be uniquely 
distinguished from each other by their isospin ($I$), isospin projection 
($I_z$) and electric charge ($q$).

\section*{References}

\end{document}